\documentclass[amsmath,floatfix,twocolumn,superscriptaddress]{revtex4-1}
\usepackage{subfigure}
\usepackage{siunitx}
\usepackage{amssymb}
\usepackage{amsmath}
\usepackage{graphicx}
\usepackage{array}
\usepackage{dcolumn}
\usepackage{psfrag}
\usepackage{bm}
\usepackage{color}
\usepackage{multirow}

\begin{document}

\title{Thermodynamics-Informed Machine Learning for Energy Materials Discovery}

\author{Pol Benítez}
\email{pol.benitez@upc.edu}
    \affiliation{Department of Physics, Universitat Politècnica de Catalunya, 08019 Barcelona, Spain}
    \affiliation{Research Center in Multiscale Science and Engineering, Universitat Politècnica de Catalunya,
    08019 Barcelona, Spain}

\author{Cibrán López}
\email{cibran.lopez@upc.edu}
    \affiliation{Department of Physics, Universitat Politècnica de Catalunya, 08019 Barcelona, Spain}
    \affiliation{Research Center in Multiscale Science and Engineering, Universitat Politècnica de Catalunya,
    08019 Barcelona, Spain}

\author{Claudio Cazorla}
\email{claudio.cazorla@upc.edu}
    \affiliation{Department of Physics, Universitat Politècnica de Catalunya, 08019 Barcelona, Spain}
    \affiliation{Research Center in Multiscale Science and Engineering, Universitat Politècnica de Catalunya,
    08019 Barcelona, Spain}
    \affiliation{Institució Catalana de Recerca i Estudis Avançats (ICREA), Passeig Lluís Companys 23, 08010 Barcelona, Spain}

\begin{abstract}
Machine learning (ML) is transforming materials discovery by enabling rapid prediction of properties that previously required computationally expensive first-principles calculations. Yet most current ML models remain fundamentally limited to zero-temperature descriptions, learning static lattice energies while neglecting the thermodynamic effects that govern materials behaviour at finite temperature. Because phase stability, functional response, and performance are governed by free-energy landscapes rather than static energies alone, this limitation represents a major barrier to predictive materials design under realistic operating conditions. In this Perspective, we argue that developing thermodynamics-informed ML constitutes one of the most important and least explored frontiers in materials discovery. We examine the fundamental shortcomings of energy-based models, highlighting the essential roles of entropy and anharmonicity in determining free energies and materials functionality. We review emerging strategies, including machine-learned interatomic potentials and hybrid ML–statistical mechanics frameworks, while identifying key challenges related to data availability, transferability, and thermodynamic consistency. Building on these advances, we outline a roadmap for thermodynamics-informed ML centred on direct free-energy learning, entropy-aware representations, and adaptive sampling across temperature. We highlight the transformative opportunities this paradigm offers for energy materials and argue that the next generation of ML models must move beyond static energy predictions towards a thermodynamic description of materials behaviour under realistic operating conditions.
\\


\end{abstract}

\maketitle

\section{Introduction}
\label{sec1}
Machine learning (ML) is rapidly transforming materials science by enabling fast and accurate prediction of properties that traditionally require computationally demanding first-principles methods. Over the past decade, ML models have achieved remarkable success in predicting energies, electronic properties, and mechanical responses across diverse classes of materials, fuelling the vision of data-driven materials discovery in which large-scale screening and optimisation can be performed with unprecedented efficiency \cite{Schmidt2019,Merchant2023}.

Despite this progress, a fundamental limitation remains largely unaddressed: most ML models are inherently restricted to zero-temperature descriptions. By construction, they are trained on datasets generated from static lattice calculations, typically within density functional theory (DFT), and therefore capture only the internal energy of materials at zero temperature \cite{Jain2013,Kirklin2015,Chanussot2021,Draxl2019}. While convenient, this approximation neglects the thermodynamic contributions that govern materials behaviour under realistic operating conditions.

\begin{figure*}
    \centering
    \includegraphics[width=1\linewidth]{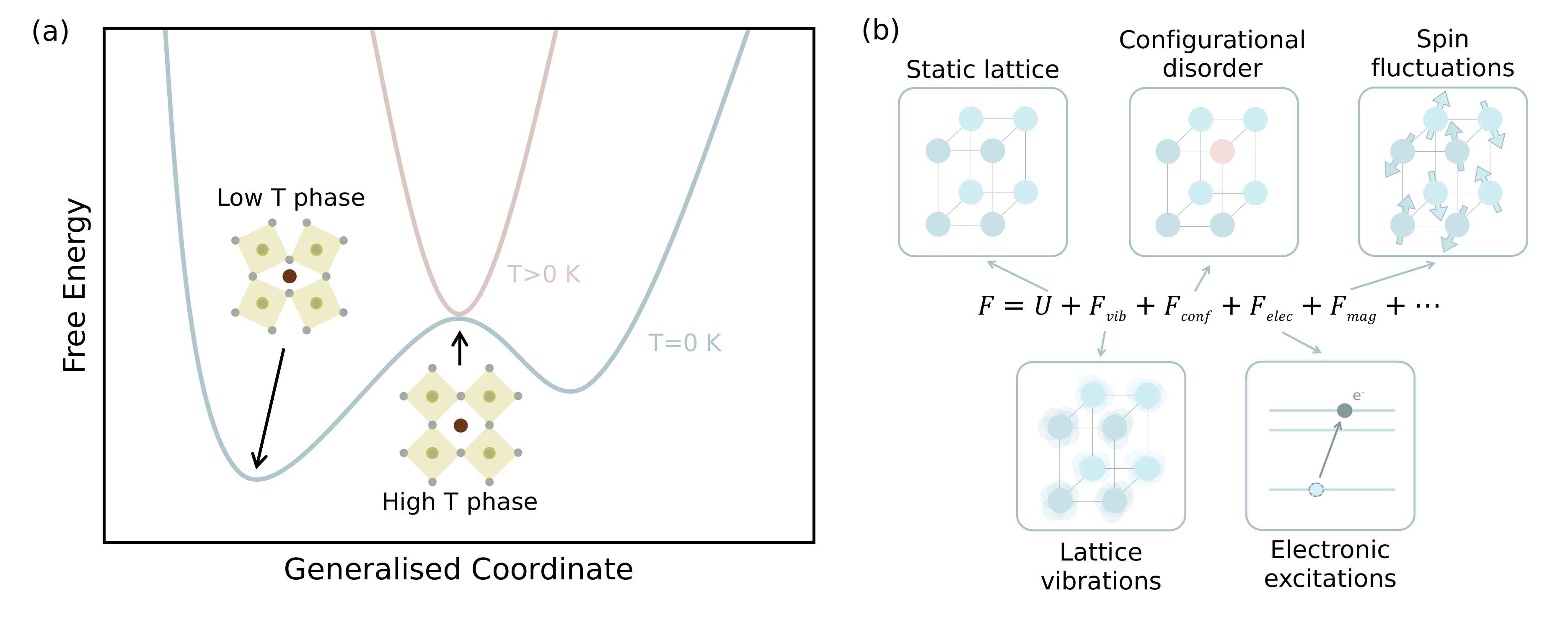}
    \caption{\textbf{Temperature reshapes the free-energy landscape of materials.} (a)~Free energy along a generalised coordinate at $T = 0$ K and $T > 0$ K. The high-symmetry phase lies in a saddle point at zero temperature but shifts to a local minimum at finite temperature, driving a transition absent from the static description. (b)~Decomposition of the free energy into static, vibrational, configurational, electronic, and magnetic contributions; conventional ML models typically learn only the first term.}
    \label{fig:fig1}
\end{figure*}

In practice, temperature plays a central role in determining phase stability, structural transformations, and functional properties (Fig.~\ref{fig:fig1}a). Entropic effects arising from lattice vibrations, configurational disorder, electronic excitations, and magnetic spin fluctuations can qualitatively alter free-energy landscapes, stabilise competing phases, and drive phase transitions that are entirely absent at zero temperature \cite{VanDeWalle2002,Widom2017,Mermin1965,Ruban2004,Cazorla2017b,Cazorla2022}. As a result, models trained solely on static energies often fail to reproduce experimentally observed behaviour, particularly in systems with strong anharmonicity and energetically competitive polymorphs.

Bridging this gap calls for a shift in perspective: from predicting energies to modelling free energies, and from static structures to thermally activated ensembles, the central premise of thermodynamics-informed ML. Machine-learned interatomic potentials have evolved from early descriptor-based schemes \cite{Behler2007,Bartok2010} to equivariant architectures and, most recently, foundational models with broad transferability across chemical space \cite{choudhary2023unified,batatia2022mace,batatia2022mace}, providing access to finite-temperature molecular dynamics trajectories from which thermodynamic quantities can be extracted via thermodynamic integration or free-energy perturbation \cite{Cazorla2012,Alfe2002,Vocadlo2002}. On a complementary front, moment tensor potentials and equivariant architectures have been combined with self-consistent phonon methods to capture anharmonic vibrational free energies beyond the harmonic approximation \cite{Eriksson2019}. Despite these advances, the incorporation of temperature as an explicit, learnable variable in ML models, rather than a \textit{post-hoc} quantity derived from expensive sampling, remains largely unsolved.

In this Perspective, we argue that developing thermodynamics-informed ML constitutes one of the most important and least explored frontiers in materials discovery. We examine the challenges and opportunities associated with incorporating finite-temperature effects into ML frameworks for energy materials discovery, and outline a roadmap centred on direct free-energy learning, entropy-aware representations, and adaptive sampling across temperature. Achieving this goal, will require not only methodological advances, but a deeper integration of ML with thermodynamic principles.

\section{Why Temperature Matters} 
\label{sec2}
Temperature is not merely an external parameter that shifts material properties by a small perturbation; it is a thermodynamic variable that can fundamentally alter the stability, structure, and functionality of a material. To appreciate why this poses a deep challenge for current ML approaches, it is useful to first recall the physical foundations that govern materials behaviour at finite temperature. We then turn to the specific ways in which the zero-temperature approximation, ubiquitous in existing ML frameworks, leads to qualitative failures in predicting experimentally observed behaviour.

\subsection{Thermodynamic foundations}
\label{sec2-1}
At the heart of materials stability and behaviour lies not the internal energy $U$, but the Gibbs free energy $G$, which governs equilibrium under realistic conditions of temperature $T$ and pressure $P$:
\begin{equation}
    G(T, P) = U - TS + PV = F + PV~,
\label{eq1}
\end{equation}
where $V$ is the volume, $S$ the entropy, and $F = U -TS$ the Helmholtz free energy. The stable phase of a material at given $(T, P)$ conditions is that which minimises $G$, not $U$. (At zero pressure, $G = F$; hence, throughout the text we will generally refer simply to the free energy.) Consequently, a phase that is energetically unfavourable at zero temperature may nonetheless become thermodynamically stable at finite temperature if it possesses a sufficiently large entropic contribution, which lowers $G$ (Eq.~(\ref{eq1})). This interplay between energy and entropy is not a secondary correction; it is the primary driver of phase transitions, order-disorder transformations, and structural polymorphism across virtually all classes of functional materials.

The entropy itself, $S = -\partial F / \partial T$, receives contributions from several distinct physical mechanisms (Fig.~\ref{fig:fig1}b):
\begin{equation}
    S = S_{\mathrm{vib}} + S_{\mathrm{conf}} + S_{\mathrm{elec}} 
        + S_{\mathrm{mag}} + \cdots,
\label{eq2}
\end{equation}
where $S_{\mathrm{vib}}$ arises from lattice vibrations (phonons), 
$S_{\mathrm{conf}}$ from configurational disorder over atomic sites, 
$S_{\mathrm{elec}}$ from thermal excitations of electrons near the 
Fermi level, and $S_{\mathrm{mag}}$ from spin fluctuations in magnetic 
systems. (Additional contributions may arise depending on the material \cite{ares25}, hence the ``$\cdots$'' in Eq.~(\ref{eq2}).) In many technologically relevant materials, including high-entropy alloys \cite{VanDeWalle2002,Widom2017}, hybrid perovskites \cite{mapi1,mapi2}, ionic conductors \cite{sagotra17,sagotra18}, and thermoelectrics \cite{jie20,cortie21}, two or more of these contributions (e.g., vibrational and configurational) are of comparable magnitude and cannot be treated independently or perturbatively.

The vibrational contribution is commonly the dominant term at low temperatures and is most 
rigorously described through the temperature-derivative of the phonon free energy. Within the harmonic approximation \cite{Baroni2001,Parlinski1997,Togo2010,cazorla2017}, the vibrational Helmholtz free energy reads:
\begin{equation}
    F_{\mathrm{vib}}(T) = k_{\mathrm{B}} T \sum_{\mathbf{q},s} 
    \left[ \frac{\hbar\omega_{\mathbf{q}s}}{2k_{\mathrm{B}}T} 
    + \ln\!\left(1 - e^{-\hbar\omega_{\mathbf{q}s}/k_{\mathrm{B}}T}
    \right) \right],
\label{eq3}
\end{equation}
where $\omega_{\mathbf{q}s}$ are the phonon frequencies at wavevector 
$\mathbf{q}$ and branch $s$. However, in materials with strong lattice 
anharmonicity, such as ferroelectrics, thermoelectrics and soft-mode 
systems, the harmonic approximation fails qualitatively \cite{Benitez2025-3,benitez-prx,wang23,mayer22}. Anharmonic effects renormalise phonon frequencies with temperature, give rise to finite phonon lifetimes, and can stabilise dynamically unstable structures that would otherwise be predicted as vibrationally unstable at zero temperature. In such cases, the full temperature-dependent free energy must be treated beyond perturbation theory, for instance by considering temperature-renormalised phonon frequencies in Eq.~(\ref{eq3}) \cite{cazorla24,Tuli2023,Carreras2017}, self-consistent phonon approaches \cite{Errea2014,Paulatto2015,Bianco2017,Monacelli2021} or direct thermodynamic integration along molecular dynamics trajectories \cite{Cazorla2012,Alfe2002,Vocadlo2002}.

In addition to maximising a target property, rationally designed materials must be synthesisable and thermodynamically stable at working conditions. A key concept in the theoretical assessment of thermodynamic stability is the convex hull, which determines the feasibility of a material to decompose into secondary phases \cite{Bartel22,Sanvito24}. The convex hull represents the lower envelope of the free energy as a function of composition, defining the set of thermodynamically stable states. Structures that lie on the hull are stable, while those above it are metastable or unstable with respect to decomposition. Determining convex hulls from free energies enables the direct evaluation of phase stability of materials under realistic conditions. As temperature increases, entropic contributions modify the relative free energies of competing phases, potentially stabilising structures that are unstable at zero temperature and/or inducing phase transitions.

\subsection{Limitations of zero-temperature ML models}
\label{sec2-2}
Current ML models for materials, including graph neural networks and equivariant architectures trained on large first-principles databases, achieve remarkable accuracy in predicting static properties such as formation energies, band gaps, and elastic constants \cite{Xie2018,Chen2019,Lopez2024-2}. Yet by construction, these models are trained almost exclusively on density functional theory calculations of relaxed structures at $T = 0$~K, and therefore learn a mapping from atomic configuration to internal energy $U$ alone. From a thermodynamic standpoint, this is a fundamental, not merely a technical, limitation.

A zero-temperature energy model has no access to $S$ and therefore cannot reproduce $G(T)$. It cannot distinguish a phase that is stable at $T \neq 0$~K conditions from one that is stable only at $0$~K; it cannot predict whether a metastable polymorph will transform upon heating; and it cannot capture the softening of free-energy barriers that governs diffusion, ionic conductivity, or thermally activated switching. In practice, this means that ML-accelerated materials screening pipelines based on convex hulls constructed from $U$ values routinely misidentify stable phases, miss thermally stabilised structures, and fail to classify competing polymorphs correctly at realistic operating temperatures.

\begin{figure*}
    \centering
    \includegraphics[width=1\linewidth]{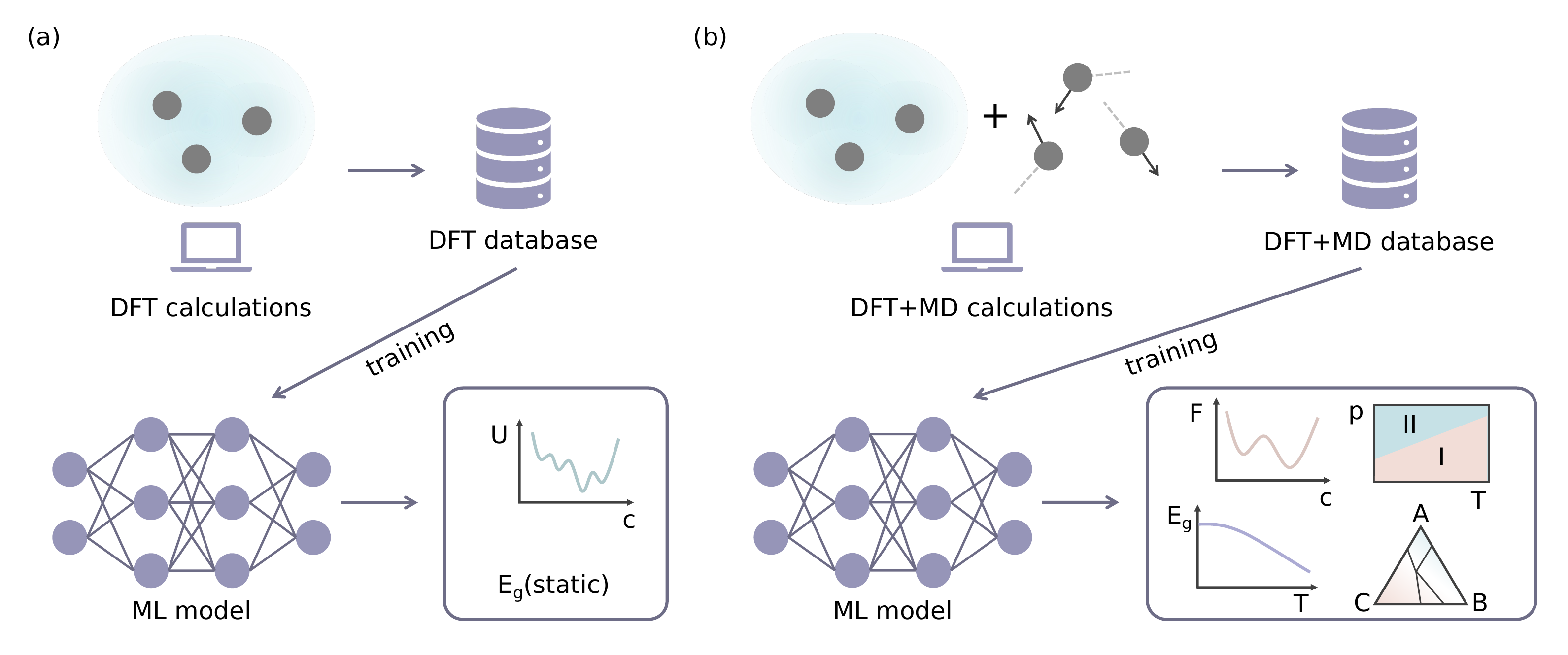}
    \caption{\textbf{From static to thermodynamic ML workflows.} (a)~Conventional pipeline: zero-temperature DFT populates a static database, and the ML model predicts the internal energy and derived quantities such as the band gap, $E_{\text{g}}$. Temperature enters nowhere. (b)~Proposed pipeline: molecular-dynamics configurations augment the training set, and the model learns $F(T)$, giving access to phase diagrams, $E_{\text{g}}(T)$, and finite-temperature convex hulls.}
    \label{fig:fig2}
\end{figure*}

This problem is made worse by the fact that many of the most technologically important materials are strongly anharmonic. In halide perovskites, for example, the high-symmetry cubic phase seen experimentally is dynamically unstable at $T = 0$~K, that is, it does not correspond to a local minimum of $U$, yet it becomes stable at room temperature due to entropic effects from anharmonic lattice fluctuations (Fig.~\ref{fig:fig1}a) \cite{Brivio2015,Patrick2015}. A model trained only on static energies will therefore either exclude this phase entirely or predict it incorrectly, no matter how accurately it reproduces DFT energetics. More broadly, this failure is not limited to perovskites: it affects any material whose free-energy landscape looks qualitatively different from its potential-energy landscape. This includes many solids with competing soft phonon modes, configurational and orientational disorder, or coupled electronic and structural behaviour. For all of these materials, current zero-temperature ML models may not be reliable.

Bridging the computational materials science gap between ideal zero-temperature and realistic finite-temperature ML predictions requires a conceptual reorientation: the target quantity must shift from the internal energy $U$ to the free energy $F$, and training data must encode thermal physics, not merely static structures (Fig.~\ref{fig:fig2}). The following sections examine strategies toward this goal and the bottlenecks that remain to be overcome.

\section{Current Strategies}
\label{sec3}
Despite the fundamental challenges outlined above, a number of ML strategies have already been developed that, to varying degrees, incorporate thermal effects into materials prediction and design. The most established of these are machine-learned interatomic potentials, which bypass the cubic scaling of \textit{ab initio} molecular dynamics by fitting a surrogate potential energy surface to first-principles data, enabling finite-temperature sampling of large supercells over extended timescales at a fraction of the computational cost. A complementary class of approaches consists of thermodynamic workflows in which ML models act as surrogates for expensive first-principles steps within established statistical-mechanics formalisms (e.g., prediction of phonon force constants). Finally, an emerging family of methods pursues the direct learning of temperature-dependent properties, treating quantities such as phonon spectra, thermal conductivity, and temperature-renormalised band structures as primary regression targets. Each of these strategies occupies a different position in the trade-off between computational efficiency, physical transparency, and the fidelity with which genuine finite-temperature thermodynamics is captured, as discussed in what follows.

\subsection{ML interatomic potentials}
\label{sec3-1}
The most direct route to finite-temperature properties is to make molecular dynamics (MD) affordable. Machine-learned interatomic potentials accomplish this by regressing the potential energy surface, and its gradients, onto a flexible functional form trained on first-principles reference data, thereby reproducing the accuracy of the underlying electronic-structure method at a computational cost several orders of magnitude lower. Once trained, an MLIP enables molecular dynamics simulations of large supercells over nanosecond timescales, from which finite-temperature thermodynamic quantities, such as free energies, heat capacities, and thermal expansion coefficients, can be extracted through thermodynamic integration, free-energy perturbation, or direct fluctuation analysis \cite{Cazorla2012,Alfe2002,Vocadlo2002}.

The field has evolved through several methodological generations. Early descriptor-based schemes, including the Behler-Parrinello neural network \cite{Behler2007}, the Gaussian approximation potential \cite{Bartok2010}, the spectral neighbour analysis potential \cite{Thompson2015}, and moment tensor potentials \cite{Shapeev2016}, encode the local atomic environment through hand-crafted symmetry functions, kernels, or polynomial invariants, and remain widely used for system-specific studies where high accuracy is required. The subsequent introduction of equivariant message-passing architectures, such as NequIP \cite{Batzner2022} and MACE \cite{batatia2022mace}, markedly improved data efficiency and accuracy by preserving the rotational symmetry of the underlying physics within the network, allowing near-first-principles fidelity from a few hundred to a few thousand reference configurations. Most recently, foundational (or universal) potentials, including MACE-MP \cite{Batatia2023} and CHGNet~\cite{Deng2023}, have been trained on vast structural databases \cite{Jain2013, Kirklin2015}, yielding broad transferability across the periodic table and enabling, in principle, finite-temperature simulations of arbitrary compositions without material-specific refitting.

Despite these advances, several limitations constrain the use of MLIP for thermodynamic prediction. Accuracy remains contingent on the diversity and representativeness of the training set: a potential fitted to configurations sampled near equilibrium may extrapolate poorly to the large-amplitude, anharmonic displacements that dominate at high temperature or near phase transitions, precisely the regime of greatest interest \cite{Subramanyam2025,Li2024}. Foundational potentials mitigate this problem through breadth of coverage, but their reliance on databases of relaxed, zero-temperature structures means that thermal and anharmonic configurations remain underrepresented, and their accuracy for soft-mode dynamics and free-energy differences between competing polymorphs is not guaranteed \cite{Liu2025}. Moreover, extracting free energies from MD trajectories still requires careful and often expensive sampling protocols \cite{Cazorla2012,Alfe2002,Vocadlo2002}. Active-learning strategies, in which configurations of high predictive uncertainty are iteratively added to the training set, have proven effective at addressing the transferability problem at manageable cost~\cite{Podryabinkin2017, Vandermause2020, Csanyi2004, Botu2017}, and are discussed further in Sec.~\ref{sec5}.

\subsection{Thermodynamic workflows}
\label{sec3-2}
A second, complementary strategy retains the established formalism of statistical mechanics and inserts ML models as surrogates for its most expensive individual steps, rather than replacing the physical framework wholesale. In this paradigm the thermodynamic machinery, whether the quasi-harmonic approximation (QHA), self-consistent phonon theory, or thermodynamic integration, remains explicit and physically transparent, while ML accelerates the first-principles evaluations on which it depends.

Within the harmonic and QHA frameworks \cite{Baroni2001,Parlinski1997,Togo2010,cazorla2017}, the vibrational free energy of Eq.~(\ref{eq3}) is obtained from the phonon spectrum, which in turn requires the second-order interatomic force constants. Computing these from density functional perturbation theory or finite displacements is tractable for simple crystals but becomes prohibitive for large or low-symmetry cells, and must be repeated at every volume to capture thermal expansion within the QHA. Replacing this step with an MLIP, or with a model trained to predict force constants directly, reduces the cost by orders of magnitude while retaining the interpretability of the phonon picture \cite{Eriksson2019, Okabe2024}. Universal potentials have been used in exactly this capacity to enable high-throughput phonon and free-energy calculations across thousands of compounds \cite{Lee2025}. The same logic extends naturally to anharmonic formalisms: moment tensor and equivariant potentials combined with self-consistent phonon or temperature-dependent effective potential methods recover anharmonic vibrational free energies beyond the harmonic approximation at a fraction of the cost of direct \textit{ab initio} sampling \cite{Shapeev2016,Monacelli2021,Hellman2013}.

A distinct advantage of this class of approaches is that thermodynamic consistency is inherited from the underlying formalism rather than learned: entropy, heat capacity, and thermal expansion follow from analytic derivatives of a physically grounded free-energy expression, ensuring that the predicted quantities obey the correct thermodynamic relations by construction. The principal limitation of these workflows is inherited from their physical scaffolding: harmonic and quasi-harmonic treatments break down for strongly anharmonic and dynamically unstable systems, while the self-consistent and integration-based alternatives that remedy this, though accelerated by ML, remain considerably more demanding and are not yet routinely applied at scale.

\subsection{Direct learning of $T$-dependent properties}
\label{sec3-3}
A third and more recent strategy dispenses with the intermediate potential energy surface altogether and treats a temperature-dependent property as the primary regression target, learning a direct map from atomic structure, and in some cases explicitly from temperature, to the observable of interest. This bypasses both the cost of MD sampling and the approximations of any particular phonon formalism, at the expense of a more opaque relationship between the prediction and its underlying physics.

The most developed instances target vibrational and transport properties. Graph neural networks have been trained to predict full phonon dispersions and densities of states directly from crystal structure \cite{Okabe2024,Al-Fahdi2025}, from which vibrational free energies and heat capacities follow, while related models regress lattice thermal conductivity for high-throughput screening of thermal materials \cite{Ojih2024,Hu2024}. A parallel line of work targets the temperature-renormalised electronic structure: rather than performing costly electron-phonon calculations, ML models learn the mapping from thermally displaced configurations, or directly from temperature, to the renormalised band gap and related optoelectronic quantities \cite{Zhong2024,Aryal2026,Benitez2025-2}. These approaches make properties available that previously required elaborate perturbative or sampling-based methods; when temperature is included as an explicit input, they also move closer to treating it as a genuine learnable variable \cite{Lopez2025}.

The characteristic weaknesses of this strategy are data scarcity and limited generalisation. Because each target property must be labelled by an expensive finite-temperature calculation, training sets are typically small and narrow in chemical and thermodynamic scope, and models trained on them extrapolate unreliably to unseen compositions or temperature ranges, particularly across phase transitions where the property may change discontinuously \cite{Lopez2025, Forslund2025}. The absence of an explicit physical intermediate also means that thermodynamic consistency is not guaranteed: a model may predict a free energy and an entropy that do not satisfy the correct differential relation unless this is imposed during training. These shortcomings, common to the direct-learning paradigm, motivate the physics-informed and multi-task frameworks discussed in Sec.~\ref{sec5}, which seek to retain the efficiency of direct prediction while restoring the thermodynamic structure that the surrogate discards.

\section{Key Bottlenecks}
\label{sec4} 
Despite the temperature-aware ML strategies already developed, as highlighted in the previous section, several key bottlenecks remain before thermodynamics can be fully incorporated into materials discovery and design. The first and most basic is data: finite-temperature datasets remain intrinsically limited by the high computational cost of sampling thermodynamic ensembles with sufficient accuracy. Capturing broad regions of configuration space typically requires expensive molecular dynamics or high-level structural calculations, making large-scale datasets impractical \cite{Herzog2024,Chang2022}. This limitation is amplified for disordered or high-entropy systems \cite{Li2024,Kaufmann2020}, where even zero-temperature datasets are already scarce, further restricting model generalisation \cite{Subramanyam2025,Frueh2025}. Indeed, recent work shows that model performance in atomistic ML depends critically on the diversity and coverage of training configurations \cite{Subramanyam2025}, reinforcing dataset quality as a foundational bottleneck on materials discovery.

This scarcity of data has direct consequences for how well models generalise across temperature. Because training sets rarely span the full range of thermally accessible configurations, ML models trained on limited thermodynamic conditions often fail when extrapolated to new temperature ranges, as the underlying configuration distribution shifts beyond what was seen during training. The problem is most acute near phase transitions \cite{Lopez2025,Forslund2025}, where abrupt structural and entropic changes occur and standard ML models, built on the assumption of smooth interpolation, struggle to capture the resulting discontinuities.

A natural way to compensate for sparse, incomplete data would be to build in physical structure that constrains predictions even outside the training domain, yet this is precisely what current architectures lack. Most ML models used in materials science do not explicitly enforce thermodynamic consistency, despite being trained on data derived from physically consistent frameworks. For instance, databases constructed from zero-temperature DFT are typically post-processed to recover finite-temperature properties, a workaround that underscores the absence of built-in thermodynamic structure in the models themselves. Without constraints linking energy, entropy, and related quantities, models may violate fundamental thermodynamic relations or produce non-physical predictions outside their training domain. This missing inductive bias is especially costly at finite temperature, where stability conditions, phase equilibria, and response functions impose strict physical requirements that current architectures rarely encode.

\section{Machine learning meets thermodynamics}
\label{sec5}
\begin{figure*}
    \centering
    \includegraphics[width=1\linewidth]{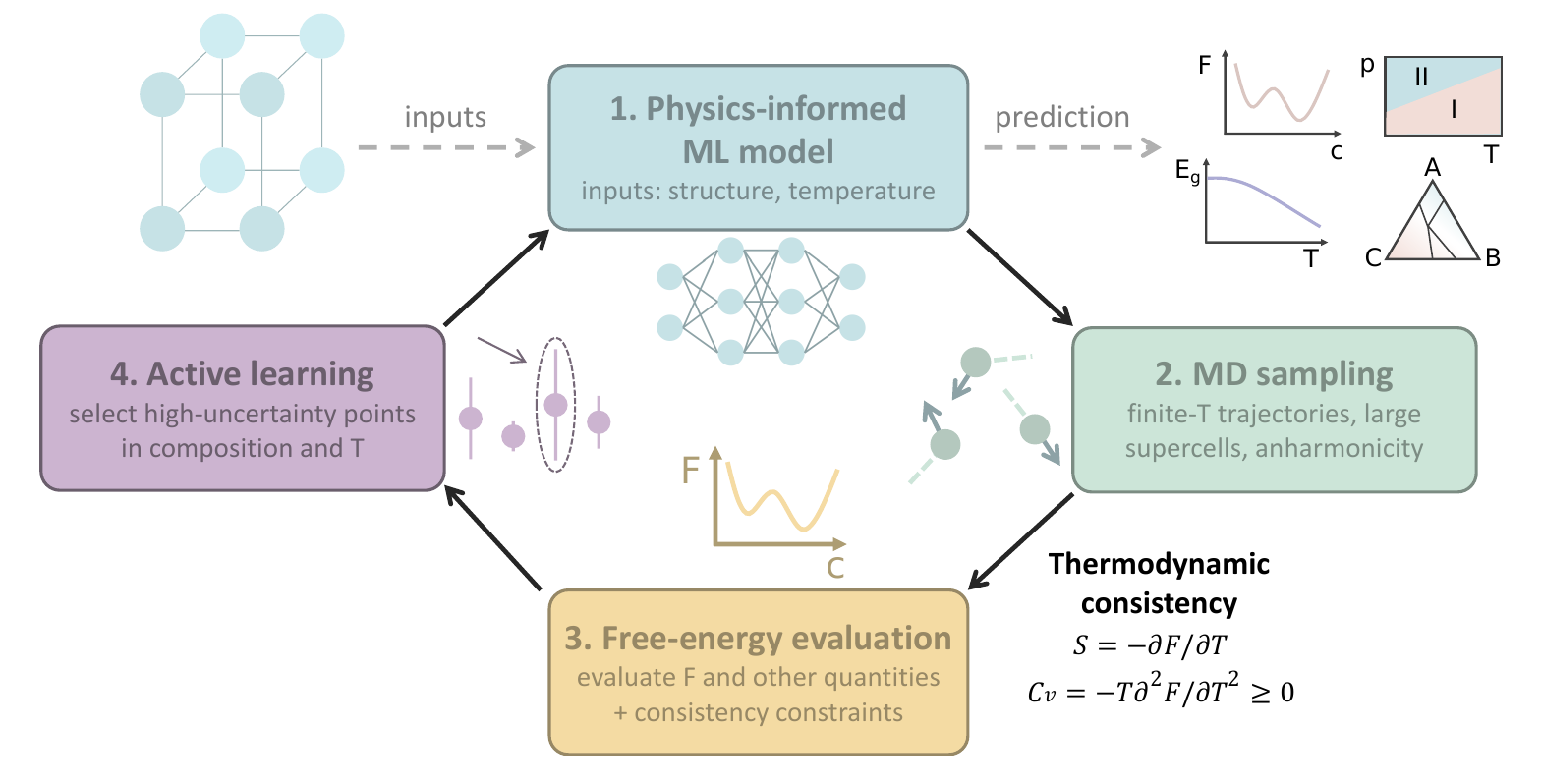}
    \caption{\textbf{Framework for thermodynamics-informed ML.} Closed loop in which temperature is an explicit input. (1)~A physics-informed model predicts $F$ from structure and temperature; (2)~MD sampling generates anharmonic finite-temperature trajectories; (3)~free energies and derived quantities are evaluated subject to thermodynamic-consistency constraints; (4)~active learning selects high-uncertainty points in composition–temperature space for high-fidelity evaluation and retraining.}
    \label{fig:fig3}
\end{figure*}

The ML-based estimation of thermodynamic quantities is inherently more complex than static energies, as they involve temperature-dependent contributions and derivatives of the free energy. However, these quantities can be naturally incorporated into ML models by embedding physical constraints and relations directly into the learning and prediction processes \cite{Hu2024,Hoffmann2026}. Figure~\ref{fig:fig3} summarises the framework we propose, in which temperature enters as an explicit input and thermodynamic consistency is enforced at each stage of a closed active-learning loop. In this context, thermodynamics-informed ML requires not only accurate predictions, but also consistency across temperature-dependent observables.

\subsection{Physics-informed ML for free energies}
\label{sec5-1}
The central question in thermodynamics-aware machine learning is not whether to include temperature effects, but how to represent them within a predictive model \cite{Lopanitsyna2021}. Several strategies have emerged, differing in what quantity is learned and how physical consistency is enforced.

A direct approach consists in learning the free energy itself as a function of temperature. This can be implemented either by conditioning the model explicitly on $T$, or by learning a parametric representation (e.g., polynomial or physics-inspired expansions) whose coefficients depend on the atomic structure \cite{Legrain2017,Deml2016,Ramprasad2017}. In practice, this corresponds to learning a joint representation of structure and temperature, which can be achieved by augmenting graph-based architectures with continuous thermodynamic variables \cite{Lopez2025,Zhang2022}. Such approaches are computationally efficient and enable rapid evaluation of thermodynamic properties across temperature ranges. Moreover, they allow one to go beyond harmonic or quasi-harmonic approximations, since no explicit assumption on lattice dynamics is required.

A key modelling challenge in this framework is ensuring that the learned free-energy surface is smooth and physically consistent \cite{Teichert2019,Laiu2022}. Architecturally, this can be addressed through the use of smooth activation functions such as SiLU and Softplus, which guarantee continuity of higher-order derivatives required for entropy and heat capacity evaluation; this consideration has motivated architectural choices in several state-of-the-art ML potentials \cite{Schutt2017,Batzner2022}. Regularisation of second- and third-order derivatives during training can further suppress unphysical oscillations, an approach analogous to the force-matching strategies used in fitting ML interatomic potentials \cite{Ercolessi1994,Chmiela2017}. 

Multi-task learning, in which the model is trained simultaneously on free energies, entropies, and heat capacities ($C_v = -T\partial^2F/\partial T^2$), reduces error amplification during temperature differentiation by enforcing thermodynamic consistency across all predicted quantities \cite{Hammad2025,Ibrahim2024}. This general strategy of learning smooth, thermodynamically consistent free-energy surfaces has been applied to predict the temperature-dependent stability of inorganic compounds and competing polymorphs \cite{Lopez2025,Forslund2025}.

An alternative strategy is to learn intermediate quantities from which the free energy is derived. The most common example is the prediction of phonon spectra or force constants \cite{Okabe2024,Al-Fahdi2025,Ojih2024,Lee2025}, from which the vibrational free-energy contribution can be estimated (Eq.~(\ref{eq3})). This approach benefits from a clear physical interpretation and strong generalisation at low temperatures, but inherits the limitations of the underlying approximations, such as the neglect of anharmonic effects in harmonic or quasi-harmonic formalisms \cite{Baroni2001,Parlinski1997,Togo2010,cazorla2017}. Similarly, ML interatomic potentials can be used to generate molecular dynamics trajectories, from which free energies are obtained via thermodynamic integration or statistical sampling, at the cost of increased computational effort \cite{Cazorla2012,Alfe2002,Vocadlo2002}.

A third class of approaches combines both direct and intermediate strategies by learning free-energy differences. A recent and technically notable example is the work by Ben-Shimon \textit{et al.} \cite{BenShimon2026}, who developed an ML framework capable of directly estimating free-energy differences between phases without requiring explicit sampling of transition pathways. The key idea is to learn the entropic contribution to the free-energy difference from short finite-temperature simulations: the model is trained on pairs of configurations drawn from distinct phases, with labels derived from thermodynamic integration over short MD trajectories. This strategy circumvents the need for long equilibration runs while retaining sensitivity to anharmonic and configurational entropy. Applied to competing polymorphs in battery cathode materials, the approach successfully ranked phase stabilities at elevated temperatures in close agreement with full thermodynamic integration, at a fraction of the computational cost \cite{BenShimon2026}. These hybrid strategies aim to retain physical interpretability while reducing computational cost.

From a modelling perspective, the approaches described in this section define a trade-off between efficiency, accuracy, and physical fidelity. Direct free-energy models offer flexibility and scalability, while indirect approaches grounded in phonons or molecular dynamics provide stronger physical constraints but may be limited in their applicability. Physics-informed ML frameworks further bridge this gap by embedding thermodynamic relations into the training process, ensuring that predicted free energies remain smooth, differentiable, and consistent across temperature.

\subsection{Learning entropic contributions}
\label{sec5-2}
The total free energy of a material arises from multiple contributions presenting distinct physical origins and thermal behaviours (Eqs.~(\ref{eq2})--(\ref{eq3})). Within this framework, entropy is not introduced independently, but is consistently obtained from the temperature derivative of each free-energy term (Sec.~\ref{sec2}). This approach naturally motivates learning individual contributions to the free energy, from which the entropy follows directly.

The entropy can also be estimated directly using approaches that exploit the formal connection between thermodynamic entropy and mutual information. A notable example is MICE (Machine-learning Iterative Calculation of Entropy) \cite{Nir2020}, which decomposes the total entropy into a sum of low-order mutual information terms estimated iteratively using k-nearest-neighbour estimators applied to configurational samples. At each iteration, the algorithm identifies the pair of subsystems that contributes the most to the residual entropy, incorporating it into the expansion until convergence. This strategy avoids the exponential scaling of exhaustive sampling methods while systematically capturing correlations beyond the mean-field. Originally demonstrated on classical spin models and jammed particle systems \cite{Nir2020}, it was recently extended by Ben-Shimon \textit{et al.} \cite{BenShimon2026} to finite-temperature molecular dynamics simulations of inorganic solids, enabling direct estimation of entropy, and hence free-energy differences between competing phases, without sampling the transition pathway.

Closely related information-theory approaches have been developed in parallel: a neural network trained on equations of state to obtain entropy as a function of density and temperature, subsequently employing it as a reaction coordinate for crystal nucleation of Lennard-Jones systems via umbrella sampling \cite{Desgranges2018}; the computable information density, a compression-based entropy metric that distinguishes ordered from disordered phases without requiring predefined structural order parameters \cite{Martiniani2019}; and a local information entropy measure serves as a general-purpose collective variable for metadynamics simulations, enabling unsupervised discovery of solid-state phase transformation pathways, including nucleation and glass formation, without prior knowledge of the reaction coordinate \cite{Li2026}.

\subsection{Active learning in temperature space}
\label{sec5-3}
Exploring the properties of materials across temperature requires navigating high-dimensional phase spaces \cite{Curtarolo2013}, where both composition and external factors may influence the free-energy landscape. In this context, machine learning must not only predict properties accurately, but also efficiently guide the discovery process toward the most relevant regions of this landscape. Active learning provides a natural framework to achieve this guidance.

Active learning is a machine learning paradigm in which the model iteratively selects the most informative data points to be evaluated with a high-fidelity method (e.g., MLIP or DFT), rather than relying on a fixed training dataset \cite{Smith2018,Lookman2019}. This strategy is particularly effective when data acquisition is expensive, as it prioritises sampling in regions of high uncertainty or maximal expected information gain. In materials science, it has been successfully applied to accelerate the exploration of chemical and configurational spaces, significantly reducing the number of required first-principles calculations while maintaining predictive accuracy \cite{Podryabinkin2017,Vandermause2020,Csanyi2004,Botu2017}.

Prior attempts to incorporate convex hull filtering into active learning loops, despite assuming zero-temperature conditions, offer instructive precedents \cite{Bisbo2022,Sanvito24,Nyshadham2019,Novick2024}. Meredig \textit{et al.} \cite{Meredig2014} pioneered the use of DFT-computed formation energies within an iterative screening workflow to identify stable ternary compounds, demonstrating that hull-based filtering could reduce the search space by orders of magnitude. More recently, frameworks such as CAMD (Computational Autonomy for Materials Discovery) \cite{Montoya2020} and related Bayesian optimisation approaches \cite{Balachandran2016} have coupled zero-temperature hull stability as a hard constraint within the acquisition function, effectively discarding thermodynamically inaccessible compositions before committing high-fidelity resources. Extending this logic to finite-temperature hulls is a natural and important generalisation (Fig.~\ref{fig:fig3}, stage 4), particularly for materials exhibiting thermally-activated ionic diffusion (e.g., solid-state electrolytes) where the relevant free-energy landscape may change substantially from absolute zero to finite-temperature operating conditions.

\subsection{Hybrid workflows}
\label{sec5-4}
Hybrid workflows that couple ML screening with selective high-fidelity verification have been demonstrated in different materials discovery contexts \cite{Kusne2020,Biswas2024}. A canonical strategy is the funnel approach: ML models first screen large candidate spaces at low cost, and DFT or beyond-DFT methods (e.g., hybrid functionals and many-body perturbation theory) are applied only on the top-ranked candidates. This has been formalised in frameworks such as Atomate \cite{Mathew2017} and AiiDA \cite{Huber2020}, which automate the handoff between ML ranking and first-principles validation within reproducible, provenance-tracked workflows. For thermodynamic stability specifically, the approach of Bartel \textit{et al.} \cite{Bartel2020} illustrates how an ML model trained on DFT formation enthalpies can pre-filter thousands of hypothetical perovskites with DFT reserved for the subset near the convex hull, thus reducing total computational cost by roughly an order of magnitude while recovering nearly all thermodynamically relevant candidates.

Human-in-the-loop strategies add a further layer of iterative refinement, particularly when ML predictions suggest unconventional or counterintuitive behaviour. The ACCELERATE framework \cite{Ling2017} formalised this paradigm by embedding domain-expert feedback directly into the Bayesian optimisation loop, allowing scientists to inject physical intuition (e.g., flagging structurally implausible candidates) that purely data-driven acquisition functions would otherwise miss. In the context of finite-temperature materials discovery, this interplay is especially valuable when competing phases differ by small free-energy margins close to experimental uncertainty, where expert judgment on the reliability of the underlying thermodynamic model remains indispensable.

\section{Opportunities in Energy Materials}
\label{sec6}
\begin{figure*}
    \centering
    \includegraphics[width=1\linewidth]{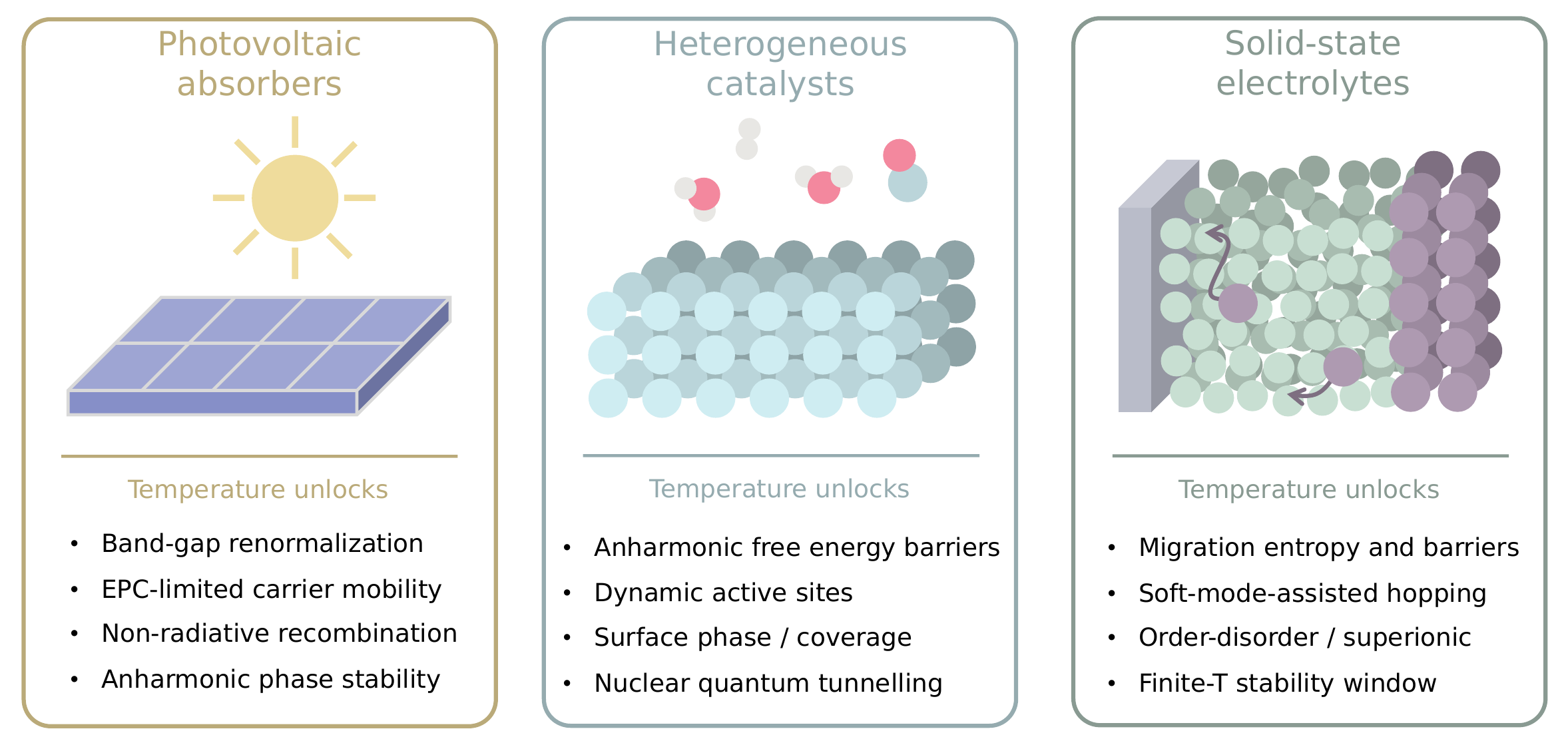}
    \caption{\textbf{Opportunities in energy materials.} Three materials families in which operating temperature is constitutive of the target property. For each family, the listed phenomena change qualitatively, not merely quantitatively, once finite-temperature effects are included.}
    \label{fig:fig4}
\end{figure*}

The rational design and theoretical prediction of energy materials, spanning photovoltaic absorbers, heterogeneous catalysts, and solid-state electrolytes (Fig.~\ref{fig:fig4}), demand a rigorous treatment that goes well beyond the static, zero-temperature picture that dominates much of contemporary computational materials science. At the heart of thermodynamic stability assessment lies the free energy convex hull: the surface in composition–structure space that separates thermodynamically stable phases from those susceptible to decomposition into competing phases or polymorphs. 

In practice, free-energy convex hulls are almost universally approximated by their zero-temperature, zero-pressure counterparts, thereby neglecting vibrational, entropic, magnetic, and anharmonic contributions that can become decisive at the operating temperatures of real devices. For materials exhibiting numerous thermodynamically metastable phases or stability windows that are narrow in temperature and composition (a common situation in halide perovskite photovoltaics, lithium superionic conductors, and multicomponent phase-change alloys), this approximation can qualitatively misrepresent the corresponding phase diagrams.

Equally critical is the ability to access temperature-renormalised collective lattice excitations, namely phonons, and their interactions with other quasiparticles as well as with particles such as electrons. Phonon dispersions and densities of states govern heat transport, thermal expansion, and mechanical response, and their accurate description at finite temperature requires accounting for anharmonic phonon–phonon interactions. Similarly, electron–phonon coupling (EPC) underpins a broad range of phenomena central to the performance of energy materials. For example, electron–phonon scattering limits carrier mobilities and governs non-radiative recombination rates in photovoltaic materials \cite{photovoltaic1,photovoltaic2}; determines the thermal behaviour and efficiency of optoelectronic materials and devices \cite{Benitez2025-2,Benitez2025-3}; mediates ionic hopping barriers and conductivity prefactors in solid electrolytes \cite{fic1,fic2}; and drives hot-carrier thermalisation and polaron formation in systems where lattice distortions are strongly coupled to charge \cite{hotpolaron1,hotpolaron2}. 

A predictive theoretical framework for energy materials must therefore incorporate free-energy methods capable of capturing, at a minimum, the full temperature dependence of phase stability, including the effects of thermally renormalized interacting phonons and anharmonicity. In the following sections, we discuss several representative families of technologically relevant energy materials, highlighting the key properties for which temperature plays a central role in determining their behaviour and thermodynamic evolution. We also review recent studies that have employed ML techniques to address conceptually and technically challenging temperature related phenomena.

\subsection{Photovoltaic materials}
\label{sec6.1}
In photovoltaic absorber materials, EPC and its temperature dependence are central to several device-critical physical processes. The most fundamental is the temperature-induced renormalisation of the electronic band gap: even at zero temperature, zero-point nuclear motion shifts the band edges through the so-called zero-point renormalisation (ZPR), and at finite temperature the gap continues to evolve in a manner that is wholly absent from static-lattice first-principles calculations \cite{Giustino2017,Gonze2020}, due mainly to EPC and the lattice thermal expansion. In strongly anharmonic and polar absorbers such as halide perovskites and chalcohalide antiperovskites, this effect can be extraordinarily large: band-gap reductions of 20--60\% relative to the zero-temperature value have been reported at room temperature, bringing theory into agreement with experiment only when EPC is properly accounted for \cite{Benitez2025-3,Benitez2025-2,Monserrat2018}. 

Beyond the band gap itself, EPC governs intrinsic carrier mobilities through electron-phonon scattering, where polar Fr\"{o}hlich coupling to longitudinal optical phonons dominates in ionic semiconductors and sets the upper bound on achievable photocurrent collection efficiency \cite{Giustino2017,Ganose2021}. EPC also underpins phonon-assisted non-radiative recombination, both at defect sites and intrinsically via multi-phonon emission, constituting one of the primary loss mechanisms limiting power conversion efficiencies below the Shockley-Queisser limit \cite{photovoltaic1,photovoltaic2}. 

From a first-principles perspective, capturing all these effects presents formidable computational challenges. Perturbative approaches to the ZPR and temperature-dependent band gap require dense Brillouin-zone sampling of electron-phonon matrix elements and break down when EPC is strong or the system is highly anharmonic, precisely the regime of the most promising emerging absorbers \cite{Gonze2020,Brousseau2023}. Computing phonon-limited carrier mobilities via DFPT combined with Wannier interpolation is accurate but demands ultra-dense reciprocal-space grids, making high-throughput screening of candidate absorbers prohibitively expensive \cite{Ganose2021}. Treating polaron formation and hot-carrier dynamics rigorously requires going beyond perturbation theory altogether, while non-radiative recombination rates via the multi-phonon mechanism necessitate real-time approaches such as time-domain \textit{ab initio} molecular dynamics, each carrying an additional layer of anharmonic and many-body complexity that remains difficult to address systematically at the first-principles level \cite{Frost2017}.

A representative body of recent ML work has made significant strides toward capturing temperature and EPC effects in the theoretical description of photovoltaic absorbers. For instance, Zhong \textit{et al.} \cite{Zhong2024} introduced HamEPC, a framework based on an equivariant graph neural network (HamGNN) trained to predict atomic orbital-based Hamiltonian matrices and their gradients, thereby bypassing the computationally prohibitive DFPT step required to obtain electron-phonon matrix elements from first principles. The framework yields EPC values in close agreement with DFPT while reducing the computational cost by several orders of magnitude; applied to GaAs along with hybrid functionals, previously inaccessible for EPC due to cost, it demonstrates that advanced electronic approaches are essential for accurate carrier mobility predictions, a quantity that is intrinsically temperature-dependent through phonon-limited scattering. 

Mosquera-Lois \textit{et al.} \cite{Mosquera-Lois2025} trained a MLIP on DFT data to model the thermodynamics of point defects in CdTe at finite temperature, systematically comparing entropic contributions from harmonic to fully anharmonic regimes via thermodynamic integration. They found that metastable defect configurations are thermally populated at room temperature and that thermal effects increase the predicted concentration of tellurium interstitials by two orders of magnitude relative to the standard zero-kelvin approximation, a striking demonstration that static-lattice defect modelling can yield qualitatively wrong predictions for operating device conditions.

Ben\'{\i}tez \textit{et al.} \cite{Benitez2025-2} addressed the question of how training data quality affects the ability of GNN models to predict finite-temperature optoelectronic properties, comparing networks trained on randomly disordered configurations against those trained on physically informed phonon displacements that selectively probe the low-energy subspace accessible to thermally excited ions. Applied to antiperovskite photovoltaic absorbers, they showed that phonon-informed datasets yield higher accuracy and robustness with significantly fewer training points, and that explainability analyses assign the greatest feature importance to chemically meaningful bonds governing band-gap variations, thereby linking predictive performance directly to the underlying electron--phonon physics. 

As another example, Aryal \textit{et al.} \cite{Aryal2026} presented a non-perturbative frozen-phonon ML scheme in which a stochastic Monte Carlo algorithm samples finite-temperature nuclear configurations from a first-principles phonon model, and a deep-learning neural network, incorporating group-theoretical symmetry-invariant descriptors, predicts the electronic property (here the temperature-dependent band gap of silicon) for each sampled configuration, achieving an order of magnitude more phonon configurations than would be feasible with direct DFT evaluation while using fewer than one hundred training calculations; this work establishes a general, non-perturbative route to vibronically renormalised band structures that is applicable to strongly anharmonic absorbers for which perturbative approaches generally  break down.

\subsection{Catalytic materials}
\label{sec6.3}
Heterogeneous catalysis is intrinsically a finite-temperature phenomenon, yet the theoretical framework most widely used to describe it, DFT combined with harmonic transition state theory (TST) and the computational hydrogen electrode, is almost entirely built upon zero-temperature, zero-pressure potential energy surfaces. The quantity governing reaction rates and selectivity at \textit{operando} conditions is the Gibbs free energy, rather than the internal energy differences between frozen geometries delivered by static DFT. Moreover, the harmonic approximation to the adsorbate partition function, universally adopted in computational catalysis, systematically underestimates vibrational and configurational free-energy contributions for weakly bound intermediates with low frustrated-translation and frustrated-rotation frequencies, and for transition states with soft modes perpendicular to the reaction coordinate, both ubiquitous at metal and oxide surfaces \cite{Nattino2019}.

At a deeper level, the catalyst surface itself is a dynamical object at reaction temperatures: surface atoms undergo large-amplitude thermal fluctuations, nanoparticles restructure, and adsorbate-induced reconstructions can qualitatively alter the active site geometry relative to the ideal zero-temperature slab universally assumed in descriptor-based approaches \cite{fan2024,Reuter2016}. This thermal disorder partially invalidates fixed-geometry adsorption energies as universal descriptors of catalytic activity, since binding energies and transition-state geometries fluctuate with temperature and a single-point descriptor at $T = 0$~K may not represent the thermally averaged quantity that actually controls turnover frequency.

Furthermore, temperature drives the system across competing catalyst phases: the stable surface termination, oxidation state, and adsorbate coverage under reaction conditions are governed by the surface free energy, rather than the zero-kelvin surface energy, and phase diagrams built from DFT total energies alone can misidentify the active phase altogether \cite{Reuter2001}. Nuclear quantum effects add a further temperature-dependent correction, particularly for reactions involving light atoms such as hydrogen, where tunnelling can enhance rates by orders of magnitude and where the crossover between tunnelling and over-barrier regimes is a phenomenon that harmonic TST is fundamentally unable to capture \cite{Richardson2016,Rommel2011}. 

Recent ML-based studies have begun to address the disconnect between static-lattice theory and \textit{operando} reality, collectively targeting the main temperature-driven failure modes of conventional zero-kelvin DFT-based catalysis. A prominent such failure concerns anharmonic corrections to free energy barriers, where static nudged-elastic-band (NEB) barriers evaluated on the zero-kelvin potential energy surface can be qualitatively incorrect. 

Schaaf \textit{et al.} \cite{Schaaf2023} introduced an active-learning protocol for GAP and MACE force fields trained on DFT data for CO$_2$ hydrogenation to methanol over indium oxide, demonstrating that finite temperature umbrella sampling not only uncovers a 40\% lower activation barrier for the previously assumed rate-limiting step, but reverses the identity of that step at operating temperatures altogether, a qualitative mechanistic revision invisible to static DFT. In a closely related study, Stocker \textit{et al.} \cite{Stocker2023} combined GAP-based MLIPs with umbrella integration to compute anharmonic free energy barriers for CHO decomposition on Rh(111), finding that the true finite-temperature barrier can be vanishingly small, whereas the harmonic TST approximation incorrectly predicts an increase with temperature, underscoring that even the sign of the thermal correction to a barrier can be wrong when anharmonicity is neglected.

A further failure mode is the assumption of static active sites: real catalyst surfaces are dynamical objects that reconstruct, sinter, and reconfigure under adsorbates at reaction temperatures, partially invalidating the use of ideal zero-kelvin slab geometries as representative models. Hou \textit{et al.} \cite{Ren2025} developed a Cu--C--O deep-potential MLIP and employed deep-potential molecular dynamics (DPMD) to reveal that, under a CO atmosphere at reaction temperatures, edge Cu atoms undergo rearrangement, ejection, and aggregation into cluster active sites in dynamic equilibrium, structures entirely absent from any static slab model, yet responsible for the observed high catalytic activity in the water-gas shift reaction. At a larger scale, Wang \textit{et al.} \cite{Jiang2025} integrated neural-network MD potentials with first-principles microkinetics within a multiscale framework for Pd-catalyzed acetylene hydrogenation, demonstrating that coupled surface-subsurface dynamics generate a distribution of Pd$_1$ single atoms and Pd$_n$ clusters whose population-weighted activity accounts for an $\sim$36,000-fold rate enhancement and $>$99\% ethylene selectivity.

Regarding the neglect of nuclear quantum tunnelling, Fang \textit{et al.} \cite{Fang2024} addressed this common oversight by developing a GPR framework for optimising semiclassical instanton pathways at surfaces, combining internal and Cartesian coordinate descriptors to handle condensed-phase periodicity. Their approach reduces \textit{ab initio} instanton optimisation to a cost comparable to that of a standard TST rate calculation, rendering quantum tunnelling corrections tractable for realistic catalytic systems. Taken together, these and other contributions demonstrate that finite-temperature, anharmonic, and quantum nuclear effects in heterogeneous catalysis are no longer beyond the reach of first-principles-quality theory, provided the computational bottleneck of direct DFT sampling can be overcome with ML techniques.

\subsection{Solid-state electrolytes}
\label{sec6.4}
In solid-state electrolytes (SSE), temperature is not a mere operational parameter but a fundamental physical variable that controls every aspect of
ionic transport, thermodynamic stability, and the very identity of the active
conducting phase. The most direct connection is through ionic conductivity
itself, which follows an Arrhenius-like dependence on temperature governed by
a migration activation energy $E_a$ and a pre-exponential factor that encodes
the attempt frequency and migration entropy. 

Commonly, these quantities are computed from zero-kelvin NEB calculations on static, perfectly ordered crystal structures. This approach neglects several physically essential contributions: the anharmonic renormalisation of the potential energy landscape along the migration pathway, the large-amplitude thermal fluctuations of the framework sublattice that dynamically open and close bottleneck geometries for ion hopping, and the configurational disorder of the mobile-ion sublattice that is often the hallmark of the superionic state itself \cite{Sagotra2019}. It is now well established that the low-energy optical phonons of the framework lattice, particularly soft modes associated with polyhedral tilts, librations, and polarisable-anion dynamics, are directly coupled to ionic mobility \cite{fic1,Muy2021}. These phonon-ion interactions are intrinsically temperature-dependent: at finite temperature, anharmonic renormalisation shifts and broadens phonon modes, and the thermally populated soft modes that facilitate concerted ion hopping are absent from any harmonic, zero-kelvin phonon calculation \cite{Gupta2021}. 

At a deeper level, many of the most technologically promising SSE (e.g., argyrodites, NASICON-type conductors, lithium thiophosphates, and closo-borane
hydrides) are thermodynamically metastable at room temperature and owe their stability and high conductivity to a temperature-driven order-disorder
transition in which the mobile-ion sublattice disorders into a high-symmetry, superionic phase with large configurational entropy. The thermodynamic stability window of these phases is therefore controlled by vibrational and configurational free energies rather than by zero-kelvin total energies, and phase diagrams constructed from static DFT can misidentify the stable phase and incorrectly predict decomposition \cite{Barroso2023,He2021}.

From a first-principles perspective, accurately capturing these temperature-driven effects presents formidable computational challenges. The gold-standard approach, \textit{ab initio} molecular dynamics (AIMD) \cite{fic1,fic2}, is in principle capable of sampling the finite-temperature potential energy landscape, anharmonic lattice fluctuations, and diffusive ion trajectories simultaneously, but its cost scales as $\mathcal{O}(N^3)$ with system size and limits accessible timescales to tens of picoseconds, often insufficient to converge diffusion coefficients at or below room temperature where ion hopping is rare. Computing $E_a$ via NEB at zero kelvin provides a cheap alternative but misses all anharmonic, entropic, and dynamical corrections to the migration free energy. Anharmonic contributions to migration barriers have been shown to reduce $E_a$ by up to $\sim$1~eV between 0~K and operating temperature in some materials \cite{fic1,Barroso2023}, demonstrating that the harmonic approximation is not merely quantitatively inaccurate but can be qualitatively misleading. 

Constructing the finite-temperature free energy convex hull required to assess thermodynamic stability against phase decomposition requires, in turn, the vibrational free energy of every competing phase, a task that demands anharmonic phonon calculations, such as those based on the self-consistent harmonic approximation (SCHA) \cite{Monacelli2021} or temperature
dependent effective potential (TDEP) methods \cite{Hellman2013}, whose cost per phase is comparable to AIMD simulations. For disordered or partially disordered SSE, one must additionally enumerate configurational microstates and sum their statistical weights, a combinatorial problem typically addressed through cluster expansions or special quasi-random structures that themselves require extensive DFT training data \cite{He2021}. 

A representative set of recent ML contributions has made substantial progress against the specific computational bottlenecks that prevent first-principles
methods from delivering a reliable, finite-temperature description of solid-state electrolytes. At the level of ionic transport, Zhou \textit{et al.} \cite{Zhou2025} developed moment tensor potentials (MTPs) for several oxide ionic conductors using passive and active learning on AIMD trajectories, enabling NVT molecular dynamics simulations on supercells of over 1,000 atoms for 10~ns timescales, that is, more than two orders of magnitude beyond what AIMD permits. With this approach, they accurately reproduced experimental diffusion coefficients, Arrhenius activation energies, and oxide-ion and proton conductivities, including the highly anisotropic, layered transport pathways that are only resolved through long-trajectory, large-scale sampling. 

Maevskiy \textit{et al.} \cite{Maevskiy2025} approached the conductivity prediction problem from a fundamentally different angle: rather than running finite-temperature MD explicitly, they demonstrated that the topology of the universal MLIP potential energy landscape itself, encoded in heuristic structural descriptors derived from GNN-evaluated energy profiles along candidate migration pathways, constitutes a reliable predictor of ionic conductivity. A ranking of Materials Project structures was performed for which 8 of the 10 highest-scoring candidates were confirmed to be superionic, at a computational cost roughly 50 times lower than ML-potential MD and at least three orders of magnitude lower than AIMD. 

Beyond transport, other contributions address the complementary problems of finite-temperature thermodynamic stability and spectroscopic characterisation. Tolborg and Walsh \cite{Tolborg2023} devised a low-cost method for incorporating vibrational free energy into cluster expansions of solid solutions by fitting an MLIP to the DFT relaxation trajectories already generated during cluster expansion construction, requiring no additional DFT calculations beyond those already performed. It was demonstrated for several ternary compounds that including vibrational entropy produces significantly better agreement with experimental miscibility gaps. This approach is directly transferable to the mixed-conductor solid electrolyte systems where configurational and vibrational entropy jointly govern the finite-temperature stability against phase decomposition, a quantity that zero-kelvin convex hull calculations entirely miss. 

Grumet \textit{et al.} \cite{Grumet2026} addressed the challenge of connecting
finite-temperature atomistic simulations to experimental spectroscopic observables: they combined an MLIP with a tensorial ML model for Raman polarisability to simulate vibrational spectra of strongly disordered superionic conductors at near \textit{ab initio} accuracy, at a fraction of the cost of DFT-based approaches. It was shown that liquid-like ionic motion dynamically lifts Raman selection rules to produce characteristic low frequency diffusive scattering, establishing this spectral feature as a practical, ML-computable descriptor for high-throughput screening of fast-ion conductors. 

As another example, Du \textit{et al.} \cite{Liu2025} performed a systematic benchmark of 12 universal MLIPs, including GRACE, DPA, MatterSim, MACE, SevenNet, and CHGNet, across energies, forces, phonons, elastic moduli, thermodynamic properties, and Li$^+$ diffusivity in solid-state electrolytes. GRACE-2L-OAM, MACE-MPA, MatterSim, DPA-3.1-3M, and SevenNet-MF-ompa were consistently identified as superior. This benchmarking work fills a critical gap by providing the community with evidence-based model selection criteria for the temperature-dependent transport and stability calculations that now underpin computational SSE discovery.

\section{Outlook}
\label{sec7}
The central message of this Perspective is that the zero-temperature approximation, while computationally convenient, represents a fundamental limitation of current ML frameworks for materials discovery rather than a minor technical shortcoming. As we have argued throughout, temperature is not an external perturbation on a fixed energy landscape but a thermodynamic variable that reshapes free-energy surfaces, stabilises new phases, and drives the functional phenomena, such as ionic conductivity and catalytic activity, that ultimately determine whether a material is useful under realistic operating conditions. 

The field has made impressive strides: machine-learned interatomic potentials now enable molecular dynamics at near first-principles quality across thousands of atoms and nanosecond timescales; foundational models provide broad transferability across chemical space; and emerging strategies for direct free-energy learning, entropy-aware modelling, and active sampling in temperature space are beginning to close the conceptual gap between static-lattice ML and genuine thermodynamic description. Yet the core challenge, incorporating temperature as an explicit, learnable variable rather than a post-hoc quantity derived from expensive sampling, remains largely unsolved.

Several specific bottlenecks have crystallised from this analysis. Training data for finite-temperature properties remains scarce and expensive to generate, particularly for disordered and high-entropy systems where even zero-temperature datasets are limited. Entropy, the quantity that governs phase stability and functional response, is not a direct observable and must be inferred through statistical sampling or thermodynamic integration, making its incorporation into ML models structurally more difficult than energy prediction. ML models trained on narrow thermodynamic windows extrapolate poorly, especially near phase transitions where the underlying configuration distribution may change discontinuously. And the absence of thermodynamic consistency constraints in most current architectures allows models to produce physically unrealisable predictions outside the training domain.
Looking forward, we identify several research directions that appear both technically tractable and scientifically relevant.

\textit{Direct free-energy learning with thermodynamic constraints.}~The most natural long-term goal is the construction of ML models that learn the free energy as a primary output, rather than deriving it as a post-hoc quantity. Progress here will require architectures that encode thermodynamic self-consistency by design, guaranteeing, for instance, that entropy is recovered as the exact temperature derivative of the learned free energy, and that heat capacity is positive. Multi-task training on free energies, entropies, and heat capacities simultaneously represents a practical near-term step, while physics-informed neural networks that embed thermodynamic relations as hard constraints offer a longer-term route to rigorous finite-temperature models.

\textit{Temperature-aware foundational models.}~The success of foundational interatomic potentials in transferring across chemical space suggests that a similar paradigm shift may be achievable for thermodynamic properties. A foundational model trained not only on static energies but also on finite-temperature data such as phonon frequencies, anharmonic force constants, and thermodynamic integration results across a broad range of compositions, could enable rapid, transferable estimation of phase stability, vibrational entropy, and temperature-dependent response functions without material-specific fine-tuning. Building the training infrastructure for such a model, including the assembly of large-scale, thermodynamically consistent finite-temperature databases, is a pressing community challenge.

\textit{Finite-temperature convex hull screening.}~The convex hull is the workhorse of computational materials discovery, yet it is almost universally approximated at zero temperature, leading to systematic misidentification of stable phases in the many material families where entropic contributions are decisive. Extending active learning workflows to operate on free-energy convex hulls, where the acquisition function accounts for temperature-dependent phase stability and the uncertainty of entropy estimates, not merely formation energies, would bring computational screening into closer contact with the experimental reality of synthesis and operation at finite temperature. 

\textit{Multiscale and multi-physics coupling.}~Many of the most technologically important temperature-driven phenomena (e.g., hot-carrier thermalisation in photovoltaics, polaron-mediated ion hopping in solid electrolytes, adsorbate-induced surface reconstruction in catalysts) involve the coupling of electronic, vibrational, and configurational degrees of freedom on timescales that span many orders of magnitude. Addressing these phenomena will require ML frameworks that operate simultaneously at multiple scales: from atomic-resolution force fields that capture anharmonic lattice dynamics, to coarse-grained models that propagate thermodynamic quantities to device-relevant length and time scales. The integration of ML electron–phonon coupling models with finite-temperature structural sampling and non-equilibrium transport theory represents a particularly promising frontier for energy materials discovery.

\textit{Uncertainty quantification and experimental feedback.}~As ML models are deployed in closed-loop discovery workflows, reliable uncertainty quantification becomes essential, both to guide active learning toward the most informative regions of temperature and composition space, and to flag predictions that lie outside the thermodynamic regimes represented in training data. Bayesian neural networks, ensemble methods, and conformal prediction approaches all merit systematic evaluation in the finite-temperature context. Equally important is tighter integration between computational predictions and experimental observables: spectroscopic signatures of anharmonicity (Raman linewidths, inelastic neutron scattering), calorimetric measurements of heat capacity and transition enthalpies, and \textit{in-situ} diffraction data on temperature-induced structural changes provide thermodynamic ground truth that can directly supervise and validate free-energy models.

Photovoltaic absorbers, heterogeneous catalysts, and solid-state electrolytes exemplify a broad class of materials in which operating temperature is not incidental but constitutive of the desired property. The convergence of large-scale thermodynamic databases, equivariant graph neural network architectures, and principled active learning strategies positions the field to address this challenge in a systematic way for the first time. Realising this potential will demand closer collaboration between the ML and statistical mechanics communities, a willingness to prioritise thermodynamic consistency alongside predictive accuracy, and sustained investment in the generation of high-quality finite-temperature training data. The reward, ML models capable of predicting not merely what a material is at absolute zero, but what it does at the temperatures where it must actually work, would represent a qualitative advance in the ambition and reliability of computational materials design.
\\

\section*{Acknowledgments}
P.B. acknowledges support from the predoctoral program AGAUR-FI ajuts (2024 FI-1 00070) Joan Oró, which is backed by the Secretariat of Universities and Research of the Department of Research and Universities of the Generalitat of Catalonia, as well as the European Social Plus Fund. C.L. acknowledges support from the Spanish Ministry of Science, Innovation and Universities under an FPU grant. C.C. acknowledges support by MICIN/AEI/10.13039/501100011033 and ERDF/EU under the grants CNS2025-165467,
PID2023-146623NB-I00 and PID2023-147469NB-C21 and by the Generalitat de Catalunya under the grants 2021SGR-00343, 2021SGR-01519 and 2021SGR-01411. This work is part of the Maria de Maeztu Units of Excellence Programme CEX2023-001300-M funded by MCIN/AEI (10.13039/501100011033).
\\

\bibliographystyle{unsrt}

\end{document}